\documentclass[11pt]{article}

\usepackage{graphicx}
\usepackage{graphicx}
\usepackage{amssymb}
\usepackage{epsfig}
\usepackage{amsgen}
\usepackage{amsmath}
\usepackage{amsgen}
\usepackage{amsmath}
\usepackage{amsfonts}

\newcommand{\BABARPubYear}    {01}

\newcommand{\BABARConfNumber} {17}

\newcommand{\LANLNumber} {0107059}

\input pubboard/babarsym

\def\fpm   {\ensuremath{f_{+-}}}
\def\fzz   {\ensuremath{f_{00}}}
\def\ratio {\ensuremath{(b_+^2 \fpm)/ (b_0^2 \fzz)}}

\def\reb   {\ensuremath{\rm{Re}(\varepsilon_B)}}
\def\eb    {\ensuremath{\varepsilon_B}}
\def\db    {\ensuremath{\delta_B}}

\def\bOne  {\ensuremath{|B^0_1\rangle}}
\def\bTwo  {\ensuremath{|B^0_2\rangle}}
\def\bL    {\ensuremath{|B^0_L\rangle}}
\def\bH    {\ensuremath{|B^0_H\rangle}}

\def\BzBz   {\ensuremath{\Bz {\kern -0.16em \Bz}}\xspace}
\def\BzbBzb {\ensuremath{\Bzb {\kern -0.16em \Bzb}}\xspace}

\def\etp    {\ensuremath{\varepsilon^+_{track}}}
\def\etn    {\ensuremath{\varepsilon^-_{track}}}
\def\epp    {\ensuremath{\varepsilon^+_{pid}}}
\def\epn    {\ensuremath{\varepsilon^-_{pid}}}
\def\npp    {\ensuremath{\eta^+_{pid}}}
\def\npn    {\ensuremath{\eta^-_{pid}}}

\def\etpm   {\ensuremath{\varepsilon^\pm_{track}}}
\def\eppm   {\ensuremath{\varepsilon^\pm_{pid}}}
\def\nppm   {\ensuremath{\eta^\pm_{pid}}}

\def\T      {\ensuremath{T}\xspace}

\long\def\inst#1{\par\nobreak\kern 4pt\nobreak
    {\it #1}\par\vskip 10pt plus 3pt minus 3pt}

\begin{document}
{\pagestyle{empty}

\begin{flushright}
\babar-CONF-\BABARPubYear/\BABARConfNumber \\
hep-ex/\LANLNumber \\
July 2001 \\
\end{flushright}

\par\vskip 1.5cm

\begin{center}
{\LARGE \bf \boldmath
Study of \T and \CP violation in \BzBzb mixing\\
\vspace{0.3cm}
 with inclusive dilepton events}
\end{center}
\bigskip

\begin{center}
\large The \babar\ Collaboration\\
\mbox{ }\\
July 20, 2001
\end{center}
\bigskip \bigskip

\begin{abstract}
We report a study of \T and \CP violation in \BzBzb mixing using an
inclusive dilepton sample collected by the \babar\ experiment at the
\pep2\ \BF. The asymmetry between 
$\ellp\ellp$ and $\ellm\ellm$ 
allows us to compare the probabilities for $\Bzb \to \Bz$ and 
$\Bz \to \Bzb$ oscillations and thus probe \T and \CP invariance.

A sample of 20,381 same-sign dilepton events is selected in the 1999--2000 
data sample, corresponding to an integrated luminosity of 20.7 \invfb on 
the \FourS resonance.
We measure a same-sign dilepton asymmetry of
$A_T=(0.5\pm1.2\pm1.4)\,\%$;
for the parameter \eb\ representing \T and \CP violations
in mixing, we obtain a preliminary result of
$$
\frac{\reb}{1+|\eb|^2}= (1.2\pm2.9\pm3.6) \times 10^{-3}.
$$
\end{abstract}
\vspace{1.5cm}

\vfill
\begin{center}
Submitted to the\\ 20$^{th}$ International Symposium 
on Lepton and Photon Interactions at High Energies, \\
7/23---7/28/2001, Rome, Italy

\end{center}

\vspace{1.0cm}
\begin{center}
{\em Stanford Linear Accelerator Center, Stanford University, 
Stanford, CA 94309} \\ \vspace{0.1cm}\hrule\vspace{0.1cm}
Work supported in part by Department of Energy contract DE-AC03-76SF00515.
\end{center}
}
\newpage

\begin{center}
\small

The \babar\ Collaboration,
\bigskip

B.~Aubert,
D.~Boutigny,
J.-M.~Gaillard,
A.~Hicheur,
Y.~Karyotakis,
J.~P.~Lees,
P.~Robbe,
V.~Tisserand
\inst{Laboratoire de Physique des Particules, F-74941 Annecy-le-Vieux, France }
A.~Palano
\inst{Universit\`a di Bari, Dipartimento di Fisica and INFN, I-70126 Bari, Italy }
G.~P.~Chen,
J.~C.~Chen,
N.~D.~Qi,
G.~Rong,
P.~Wang,
Y.~S.~Zhu
\inst{Institute of High Energy Physics, Beijing 100039, China }
% ...

\end{center}\newpage

%
%

\setcounter{footnote}{0}

\section{Introduction}
\label{section:intro}
Since the first discovery of \CP violation in 1964~\cite{CP64}, the kaon system
has provided many other results probing the \CPT and \T discrete 
symmetries~\cite{CPLEAR99}.
The \babar\ experiment is not limited to an investigation of \CP violation 
through the
measurement of $\sin(2\beta)$ or the angle $\alpha$; in a similar way as for 
kaon system 
studies, 
it is also possible to investigate
\CP violation purely in mixing 
and disentangle whether this \CP violation is due to \T or \CPT violation.

In this document, we have adopted a formalism very similar to that used for 
the kaon 
system~\cite{BB99}. In the absence of a \CP phase, we can define the \CP 
eigenstates 
\bOne\ and \bTwo\ as
\begin{eqnarray*}
        \bOne & = & \frac{1}{\sqrt 2}\bigl( |\Bz\rangle + |\Bzb\rangle\bigr),\\
        \bTwo & = & \frac{1}{\sqrt 2}\bigl( |\Bz\rangle - |\Bzb\rangle\bigr).
\end{eqnarray*}
The physical states (solutions of the effective Hamiltonian) can be written as
\begin{eqnarray*}
\bL & = & \frac{1}{\sqrt {1+|\eb+\db|^2}}\bigl[ \bOne
        + (\eb+\db)\bTwo \bigr],\\
\bH & = & \frac{1}{\sqrt {1+|\eb-\db|^2}}\bigl[ \bTwo
        + (\eb-\db)\bOne \bigr].
\end{eqnarray*}
In the case of \CPT and \CP invariance, $\db$ is equal to 0. Similarly, 
\T and \CP invariance gives $\eb=0$.
This means that \CP violation in mixing requires either $\eb\neq0$ or $\db\neq0$.
Inclusive dilepton events in \babar\ provide a very large sample with
which to
study \CP violation in mixing and test \T and \CPT conservation.
The semileptonic (muon or electron) branching fraction of $B$ mesons
is about 20\%. Therefore, dilepton events represent 4\% of all \upsbb 
decays. The flavor of the $B$ is tagged by the sign of the lepton. 
Assuming \CP invariance in direct semileptonic decays,
the asymmetry between same-sign dilepton pairs, $\ellp \ellp$ and 
$\ellm \ellm$, allows a comparison of the two oscillation probabilities
$P(\Bzb \to \Bz)$ and $P(\Bz \to \Bzb)$
and therefore probes \T and \CP invariance:
\begin{equation}
A_{T}(\deltat) = \frac {P(\Bzb \to \Bz, \deltat)-P(\Bz \to \Bzb, \deltat)}
                       {P(\Bzb \to \Bz, \deltat)+P(\Bz \to \Bzb, \deltat)}
\approx \frac {4\reb}{1+|\eb|^2}.
\label{at}
\end{equation}
This asymmetry\footnote{In another 
formalism~\cite{PhysBook}, 
the physical states are related to the \Bz flavor eigenstates by 
$\bL = p|\Bz\rangle +\, q|\Bzb\rangle$ and 
$\bH = p|\Bz\rangle -\, q|\Bzb\rangle$ 
where $p$ and $q$ are complex mixing parameters with the normalization 
$|p|^2 + |q|^2 = 1$.
The charge asymmetry in terms of $|q/p|$ is therefore equal to
$A_{T}=(1-|q/p|^4)/(1+|q/p|^4)$.}
does not contain the \CPT violation parameter $\db$
to first order. For this asymmetry to be different from zero
both \T and \CP violation in mixing are required.
In the approximation that $|\Gamma_{12}|<<|M_{12}|$,
$A_T = {\rm Im}(\Gamma_{12}/M_{12})$ where 
$M_{12} - \frac{1}{2}i\Gamma_{12}$ is the off-diagonal element
of the complex effective Hamiltonian for the \Bz - \Bzb system.
Standard Model calculations~\cite{SM} predict 
the size of this asymmetry to be of order $(0.5-5.0) \times 10^{-3}$.
Therefore, a large measured value for this asymmetry could be an 
indication of new physics.

The measurement of $A_T$ reported here 
is performed using events collected by the 
\babar\ detector at the \pep2 \abf\ between October 1999 and October 2000. 
The integrated luminosity of this sample is 20.7 \invfb taken 
on the \FourS resonance (``on-resonance'') 
and 2.6 \invfb taken 40 \mev below the resonance (``off-resonance'').
The \babar\ detector and its performance are described 
elsewhere~\cite{BaBarNIM01}. 

The organization of this paper is as follows.
The particle identification and event selection are described in 
Sec.~\ref{section:evsel}. In particular, this section shows how  
cascade leptons from charm decays
from charged $B$ or unmixed neutral $B$ events are 
rejected 
with a neural network (NN) approach and a \deltaz cut at 200\mum.
The method to correct the charge asymmetry in the detection of the leptons is
explained in Sec.~\ref{section:chargeAsy}. Finally, Sec.~\ref{section:mesReB} 
shows 
the details of the fit on data and gives the evaluation of 
systematic uncertainties. To avoid the possibility of bias, both the time 
distribution of the
charge asymmetry and the number of positive and negative same-sign dileptons 
are blinded during
the analysis. The unblinding of the $A_T$ measurement is performed once all 
studies of systematic uncertainties are finished.

%
%

\section{Selection of dilepton events}
\label{section:evsel}

In this study of \T and \CP asymmetries, the flavor of the $B$ meson at 
the time of its decay is 
determined by the sign of direct leptons produced in semileptonic $B$ decays. 
This section describes the selection of leptonic tracks and the rejection of 
cascade leptons.

\subsection{Lepton identification}
Lepton candidates must have a distance of closest approach 
to the nominal beam position in the 
transverse plane less than 1\cm, a distance of closest 
approach along the beam direction less than 6\cm, at 
least 12 hits in the Drift Chamber (DCH), at least one $z$-coordinate 
hit in the Silicon Vertex Tracker (SVT), 
and a momentum in the \FourS center-of-mass system (CMS) between 
0.7 and 2.3\gevc.

Electrons are selected by specific requirements 
on the ratio of the energy deposited in the Electromagnetic Calorimeter (EMC)
and the momentum measured in the DCH, 
on the lateral shape of the energy deposition in the calorimeter, and on the 
specific 
ionization density measured in the DCH. 
Muons are identified through the energy released in the calorimeter, as 
well as 
the strip multiplicity, track continuity and penetration depth in the 
Instrumented Flux Return (IFR).
Lepton candidates are rejected if they are
consistent with a kaon or proton hypothesis according to the Cherenkov angle
measured in the Detector of Internally Reflected Cherenkov Light (DIRC) 
and to the ionization density measured in the DCH.

The performance and charge asymmetry of the lepton selection are determined 
with
data control samples, as a function of the particle momentum as well as the 
polar and 
azimuthal angles. The electron and muon selection efficiencies are about 92\% 
and 75\%, with pion misidentification probabilities around 
0.2\% and 3\%, respectively. All corrections of the charge asymmetry in 
the identification
of leptons are discussed in Sec.~\ref{section:chargeAsy}. 
 
\subsection{Background rejection}
Non-\BB events are suppressed by
requiring the Fox-Wolfram ratio of second to zeroth order moments~\cite{FW} 
to be less than 0.4. 
In addition, the residual contamination from radiative Bhabha and two-photon 
events is 
reduced by requiring the squared invariant mass of the event to be greater 
than 20 (\gevcc)$^2$, the event aplanarity to be greater than 0.01, and the 
number of
charged tracks to be greater than four.

Electrons from photon conversions are identified and rejected
with a negligible loss of efficiency for signal events. 
Leptons from \jpsi\ and $\psi (2S)$ decays are identified by pairing 
them with other oppositely-charged candidates of the same lepton species, 
selected with looser criteria. We reject the whole event if any combination 
has an invariant mass within $ 3.037<M(\ellp\ellm)< 3.137\gevcc$ or
 $ 3.646 <M(\ellp\ellm)< 3.726\gevcc$.

\subsection{ Selection of direct dileptons}

To minimize the dilution due to wrong-sign leptons from cascade charm
decays (produced in $b \to c \to \ell$ transitions), we use a NN
algorithm that combines five discriminating variables. These are calculated
in the CMS and are
\begin{itemize}
\item the momenta of the two leptons with highest momentum, $p^*_1$ and $p^*_2$;
\item the total visible energy $E_{tot}$ and the missing momentum 
$p_{miss}$ of the event; and
\item the opening angle between the leptons, $\theta_{12}$. 
\end{itemize}
\begin{figure}[hbtp]
\begin{center}
\mbox{\includegraphics[height=\textwidth]{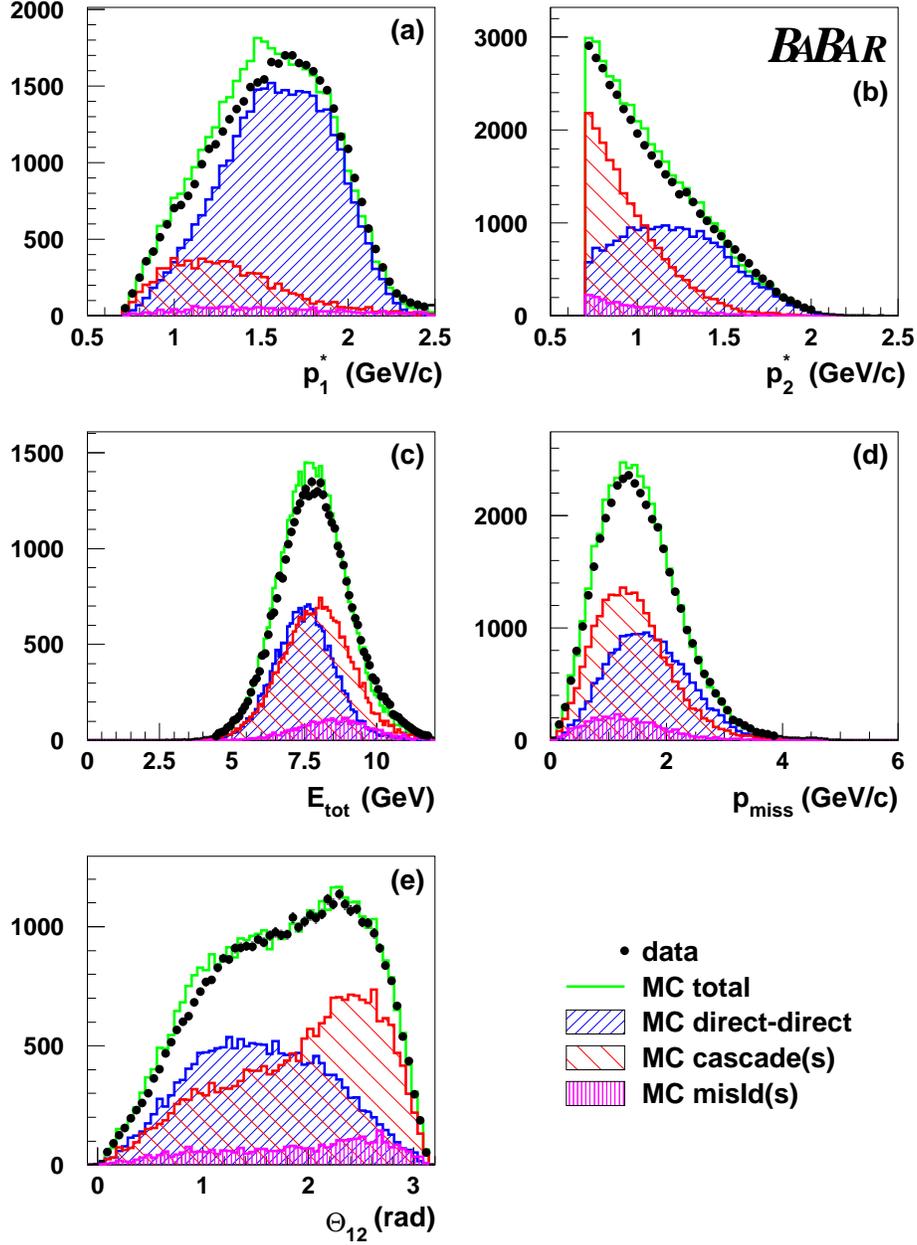}}
\end{center}
\caption{Distributions of the discriminating variables (a) $p^*_1$, 
(b) $p^*_2$, (c) $E_{tot}$, (d) $p_{miss}$ and (e) $\theta_{12}$,
for data (points) and Monte Carlo events (histograms). 
The contributions from direct-direct pairs, direct-cascade or cascade-cascade pairs, 
and pairs with one or more fake leptons are 
shown for the Monte Carlo samples.}
\label{dis_var}
\end{figure}
The distributions of these variables are shown in Fig.~\ref{dis_var},
for data and Monte Carlo simulation. 
The first two variables, $p^*_1$ and $p^*_2$, are very powerful in 
discriminating between direct and
cascade leptons. The last variable, $\theta_{12}$, efficiently removes 
direct-cascade lepton pairs coming from the same $B$ and further rejects 
photon conversions. 
Some additional discriminating power is also provided by the other two 
variables.
The NN architecture (5:5:2) consists of three layers, 
with two outputs in the last layer,
one for each lepton in the event. The network is 
trained with 40,000 dileptons from generic Monte Carlo \BzBzb and \BpBm
events. The outputs are chosen to 
be 1 and 0 for direct and cascade leptons, respectively.
The same network is used for both electron and muon selection.
We require both outputs to be greater than 0.8.

\subsection{\boldmath Background rejection based on \deltaz information}
\label{sec:DzCut}

In the inclusive approach used here, 
the $z$ coordinate of the $B$ decay point is 
the $z$ position of the point of closest approach between the lepton
candidate and 
an estimate of the \FourS decay point in the transverse plane. The
\FourS decay point
is obtained by combining the beam spot constraint and the relative position 
of the two lepton
tracks. The time difference $\deltat$ between the two $B$ mesons
is determined from the difference in $z$ between the two $B$'s by
$\deltat=\deltaz/ \langle\beta\gamma\rangle c$, where
$\langle\beta\gamma\rangle \approx 0.56$.

\begin{figure}[hbtp]
\begin{center}
\mbox{\epsfig{file=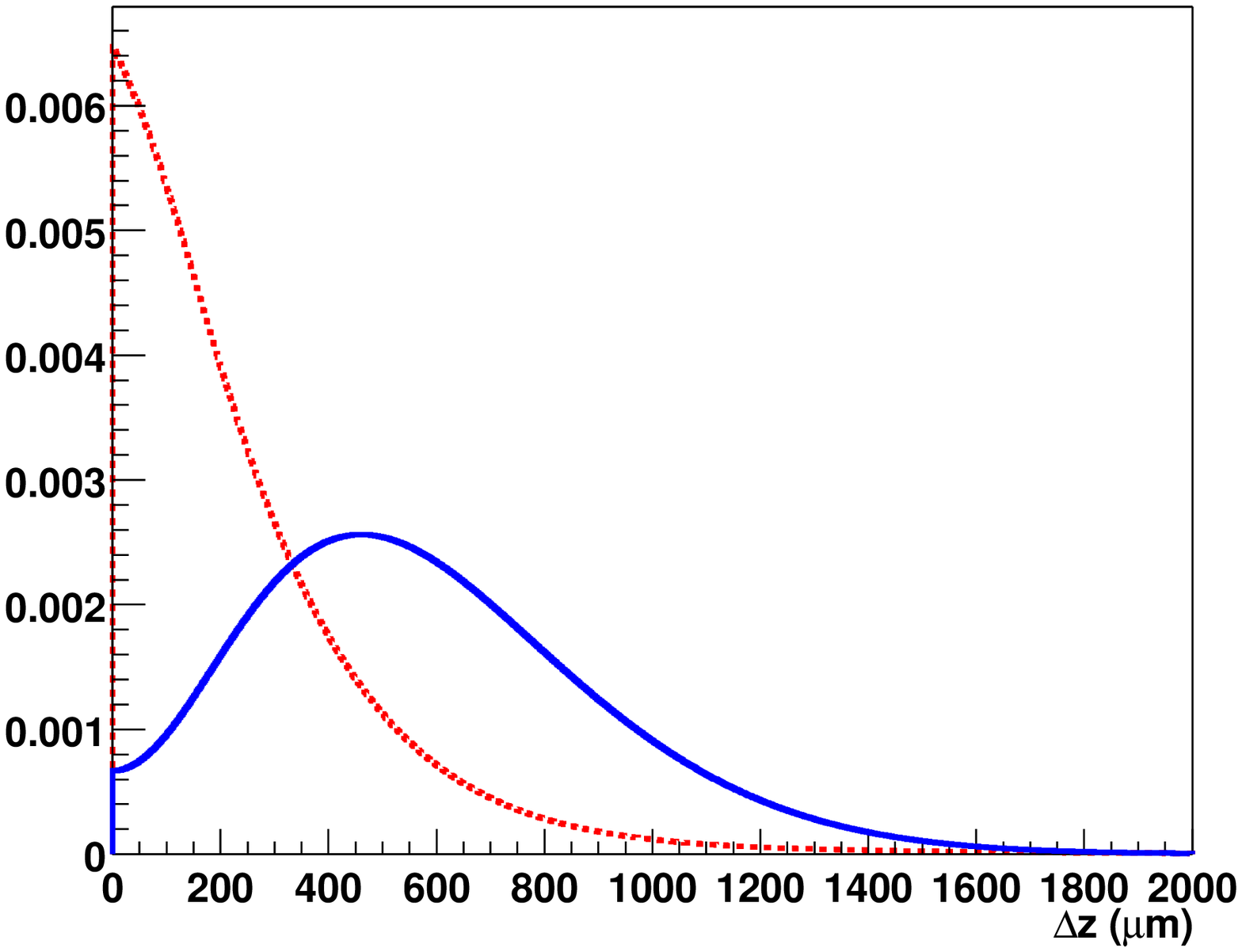,height=8.0cm}}
\end{center}
\caption{Probability density functions determined from a fit to the data 
with the value of \deltamd fixed to the world average value~\cite{PDG}, 
for the signal (\BzBz or \BzbBzb pairs) (solid line) and the background 
(cascade from unmixed \BzBzb and \BpBm events, and  
non-\BB events) (dotted line) as a function of \deltaz.}
\label{pdfSame}
\end{figure}

To measure $A_T$, we want to select mixed events 
(\BzBz or \BzbBzb pairs) whose time dependence varies as 
$e^{-\deltat/\tau_{\Bz}}\left[1-\cos(\deltamd\deltat)\right]$.
The main sources of background (cascade leptons
from unmixed \BzBzb events and \BpBm events) vary respectively as
$\sim\! e^{-\deltat/\tau_{\Bz}}\left[1+\cos(\deltamd\deltat)\right]$  
and $\sim\! e^{-\deltat/\tau_{\Bpm}}$
(see Fig.~\ref{pdfSame}). 
Therefore, a requirement of $\deltaz > 200$\mum allows us 
to eliminate 
a large fraction of background without dramatically decreasing the signal 
efficiency. 
A \deltaz cut is also effective at removing
backgrounds such as non-\BB events or \jpsi decays. Finally, 
in the measurement of $A_T$, the dilution factor due to remaining 
background will
be corrected as a function of \deltaz.

\subsection{Event yields and sample composition}
\label{section:yields}
Application of the selection criteria described above results in a
sample of 20,381 same-sign dilepton events, consisting of
5,252 electrons pairs, 5,152 muons pairs and 9,977 electron-muon pairs.

The fraction of non-\BB events, measured with the off-resonance data,
is 4.3\% with a charge asymmetry of $(-5 \pm 10)\%$.

%
%

\section{Detector-induced asymmetries}
\label{section:chargeAsy}

Since the asymmetry $A_T$ is expected to be small, we must 
control possible
charge asymmetries induced by the detection and reconstruction of
electrons and muons. The strategy in this analysis is first 
to correct on an event-by-event basis charge asymmetries determined with 
independent control samples, and then to validate this approach with a sample
of single direct leptons from $B$ decays.

\subsection{Charge asymmetry correction}

The three sources of charge asymmetry in the selection of lepton candidates are
\begin{enumerate}
 \item the difference in tracking efficiency for positive and negative 
       particles, $\etp \neq \etn$;
 \item the difference in particle identification efficiency for positive and 
       negative leptons, $\epp \neq \epn$; and
 \item the difference in  misidentification probability for positive 
       and negative particles, $\npp \neq \npn$.
\end{enumerate}
These efficiencies and probabilities are estimated using independent samples 
(see next sections) on an event-by-event basis and as a function of 
several kinematic variables $x_i$ of the charged track:
the total momentum, transverse momentum, polar angle and the azimuthal
angle of the charged track. The numbers of ``detected'' positive 
and negative leptons $N^\pm_{detected}(\ell)$ 
are related to the numbers of true leptons $N^\pm_{true}(\ell)$  by the 
equation:
\begin{equation}
\begin{aligned}
N^\pm_{detected}(\ell) & = & 
N^\pm_{true}(\ell)\cdot\etpm(x_i)\cdot\bigl[\eppm(x_i) 
+ r(\pi,p^*)\cdot\nppm(\pi,x_i)\\
&& + r(K,p^*)\cdot\nppm(K,x_i) + r(p,p^*)\cdot\nppm(p,x_i)\bigr],
\end{aligned}
\label{weight}
\end{equation}
where $r(\pi,p^*)$, $r(K,p^*)$ and $r(p,p^*)$ are the relative abundances of 
hadrons ($\pi$, $K$,
and $p$) with respect to the lepton abundance for a given $p^*$ (the momentum 
of the track in the CMS).
These quantities are obtained from generic \BB Monte Carlo events, after
applying the event selection criteria.
To correct for the charge asymmetries, 
we apply a weight proportional to the ratio 
$ N^\pm_{true}(\ell)/ N^\pm_{detected}(\ell)$, for each lepton in the sample.

\subsection{Correction with control samples}
\subsubsection{Charge asymmetry in tracking}
The event selection requires at least 12 DCH hits per track, 
which can introduce
a charge asymmetry in the tracking efficiency. 
In this analysis,
we are selecting tracks with large momentum in the CMS, which implies
that the transverse momentum is also large. Therefore, the dilepton sample 
should be only
slightly affected by any charge asymmetry in tracking. To remove this
possible bias, we determine separate efficiencies for positive and negative
particles. 
The tracking efficiency, which is dominated by the DCH, 
is defined as the ratio 
of the number of SVT tracks 
with 12 DCH hits divided by the initial number of SVT tracks. 
These track efficiency correction tables are computed as a function of 
transverse
momentum, polar angle and azimuthal angle in the laboratory frame. The charge asymmetry correction 
is less 
than 0.1\% on average in the relevant ranges for the lepton tracks.

\subsubsection{Charge asymmetry in lepton identification}
Since the particle identification efficiencies and the misidentification 
probabilities are determined
as the ratio of the number of events with an identified track over the number 
of initial events,
 the efficiency and misidentification probability tables are independent of 
the 
tracking efficiency.
These tables are computed as a function of total momentum, polar angle and 
azimuthal angle in the laboratory frame for different control samples:
\begin{itemize}
\item The identification efficiencies for the electrons are measured
with the combination of two control samples: $\gamma\gamma \to eeee$ and 
radiative Bhabha events;
\item The identification efficiencies for the muons are measured 
with a control sample consisting of $\gamma\gamma \to ee\mu\mu$ events;
\item The misidentification probabilities are determined using control
samples of kaons produced in $D^{*+}\to\pi^+D^0\to\pi^+K^-\pi^+$ decays 
(and charge conjugate), 
pions produced in $K_S\to\pipi$ decays, one-prong and three-prong $\tau$
decays, and protons
produced in $\Lambda$ decays.
\end{itemize}
For the electrons, the charge asymmetry in the efficiency reaches
(0.5--1.0)\% in some regions of the lepton phase space.
The impact of the charge asymmetry in misidentification is 
negligible 
because the absolute misidentification probability for pions is extremely 
small ($\sim 0.2\%$).
However, the $\Lambda$ control sample indicates a very large 
misidentification 
probability for antiprotons with momentum $\sim 1 \gevc$. 
Such an effect is due to the annihilation 
of antiprotons with protons in the calorimeter, which produces a signature 
similar to that of
an electron. 
The impact of this effect is balanced by the low relative abundance of 
antiprotons in generic 
$B$ events. Overall, antiprotons induce a charge asymmetry 
of order 0.1\% and a correction is applied for this effect.

For the muons, the $ee\mu\mu$ control sample shows that the charge asymmetry 
in the efficiency reaches 0.5\%. The fraction of fake pions
($\sim 3\%$) is much larger than in the case of electrons but 
there is no indication of
any charge asymmetry. On the other hand, the kaon misidentification 
distribution shows
a charge asymmetry at the level of (10--20)\% due to the difference between 
cross sections for \Kp and \Km mesons interactions
with matter in the range of momentum around 
1\gevc.

\subsection{Validation with single direct leptons}
\label{section:single}
A cross check of the correction for charge asymmetries in the lepton selection
is performed with 
an independent sample that has the same topology and kinematics as the 
leptons from dilepton events: 
namely direct leptons from semileptonic $B$ decays. The selection 
of these events is quite similar to the dilepton selection described in 
Sec.~\ref{section:evsel}.

The single-lepton charge asymmetry is sensitive to the charge asymmetry 
due to detection bias
but it may also be affected by the real physical charge asymmetry $A_T$ in the 
dilepton events. In this case, the physical charge asymmetry 
$A_{single}^{physics}$ can be written as
$$
A_{single}^{physics}= D\cdot\chi_d\cdot\frac{1}{1+R}\cdot A_T,
$$ 
where $D$ is a dilution factor due to background (cascade leptons, etc.),
$\chi_d$ is the \Bz mixing parameter, and $R$ is the fraction of 
charged $B$ mesons in the sample.\footnote{ $R$ is the ratio
$\ratio$ where $b_+$ and $b_0$ are 
respectively the semileptonic branching fractions of
charged and neutral $B$ mesons, and
$\fpm/\fzz$ is the production ratio of charged and neutral $B$ mesons.}
The possible
bias introduced by $A_T$ is suppressed by more than one order of magnitude 
and is therefore neglected.

\begin{figure}[hbtp]
\begin{center}
\mbox{\epsfig{file=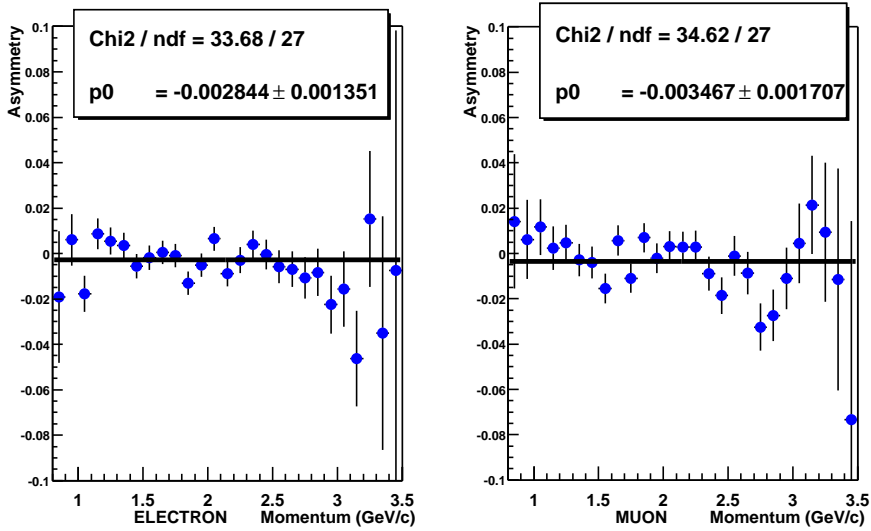,height=8.cm}}
\end{center}
\caption{Single direct lepton control sample: charge asymmetries for 
electrons (left plot)
and muons (right plot) as a function of the total momentum.}
\label{single}
\end{figure}

With the 1999--2000 data set, we select roughly 1.5 million
electrons and 1.5 million muons. 
After subtraction of scaled off-resonance data and after applying a 
correction weight as
defined in Eq.~\ref{weight}, we measure the remaining asymmetry as a 
function of
the total momentum (see Fig.~\ref{single}). For both muons and electrons,
the distributions are consistent with being flat. The single-muon charge 
asymmetry $(-0.35 \pm 0.17)\%$ 
and the single electron charge asymmetry $(-0.30 \pm 0.14)\%$ 
are consistent with zero.
These measured charge asymmetries for single leptons show
that the systematic errors related to the muon and electron detection are 
at the level of $\pm0.35\%$ and $\pm0.30\%$, respectively.


%
%

\section{\boldmath Measurement of \reb}
\label{section:mesReB}

\subsection{\boldmath Measurement of $A_T$}

\begin{figure}[hbtp]
\begin{center}
\mbox{\epsfig{file=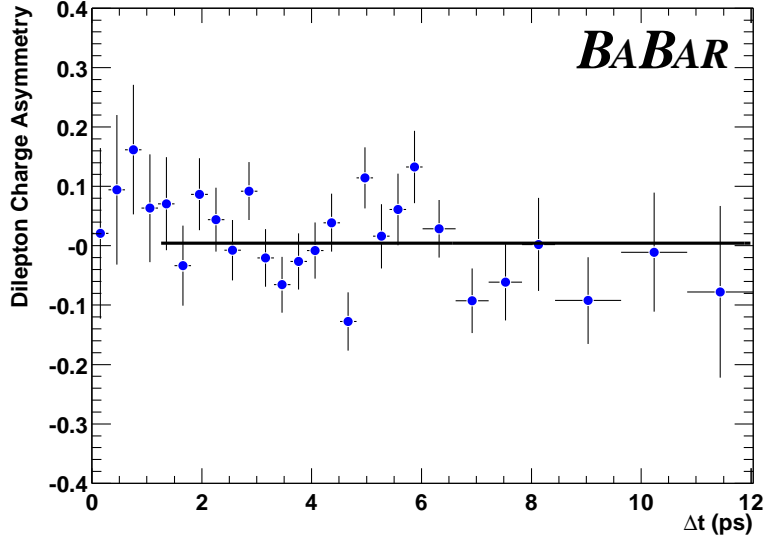,height=8cm}}
\end{center}
\caption{Corrected dilepton charge asymmetry as a function of \deltat. 
The line shows the result of the fit for the same-sign dileptons with 
 $\deltaz > 200$\mum.}
\label{atresultsWbckg}
\end{figure}

Equation~\ref{at} is applicable for  pure signal (\BzBz and \BzbBzb pairs). 
However, the dilepton
events are contaminated by cascade leptons from 
\BpBm and unmixed \BzBzb events
(see Fig.~\ref{pdfSame}), non-\BB events, and \jpsi decays.
Assuming no charge asymmetry in the 
background and assuming \CP invariance holds in the direct leptonic 
$B$ decays,\footnote{In the literature, 
the equality of the decay probabilities
$P(\Bz\ra\ellp)$ and $P(\Bzb\ra\ellm)$
is usually obtained from \CPT invariance in the decay
for inclusive production of direct leptons.
However, in this analysis, the signal is selected by imposing
cuts on the lepton momentum, which is equivalent to considering
a partial decay channel.
As a consequence, the equality of the partially integrated spectrum
for leptons and antileptons requires \CP symmetry, which is a 
stronger assumption.}
we can write
the measured asymmetry $A^{meas}_T$ (see Fig.~\ref{atresultsWbckg})
in terms of the number of events $N$ as
\begin{equation}
A^{meas}_{T}(\deltat)= 
\frac {N(\ellp\ellp,\deltat)-N(\ellm\ellm,\deltat)}
{N(\ellp\ellp,\deltat)+N(\ellm\ellm,\deltat)}
= A_{T}\cdot \frac{S(\deltat)}{S(\deltat)+B(\deltat)},
\label{atmes}
\end{equation}
\noindent 
where $S(\deltat)$ is the number of signal events and $B(\deltat)$ the total 
number of background events. 
The assumption of no charge asymmetry in the background is confirmed by the 
off-resonance data where the charge asymmetry 
$(-5\pm10)\%$ 
is consistent with zero. In addition,
the charge asymmetry of the events with $\deltaz< 100$\mum, which contain
85\% background (cascade leptons from \Bpm and unmixed \Bz),  is 
$(1.2 \pm 1.4)\%$ and so is
also consistent with zero. Finally, a possible dilution of $A_T$ due 
to double mistag is neglected because the probability of double mistag is at 
the level of only 1\%. 
All these assumptions are taken into account in the determination of the 
systematic error.

Therefore, extraction of a value for $A_T$ requires a
determination of the dilution factor\break 
$S(\deltat)/\left[S(\deltat)+B(\deltat)\right]$.
The dilution factor
can be measured directly from the data with the probability density 
functions (Fig.~\ref{pdfSame}) 
obtained from the full dilepton sample with the 
value of \deltamd fixed to the world average value~\cite{PDG}. 
With this method, the fraction of non-\BB events is determined
from off-resonance data, and the fraction of cascade leptons and the 
resolution function corrections
are measured directly from the dilepton data.
In order to benefit in the 
determination of $A_T$
from the full information provided by \deltaz, 
a correction factor for the background dilution 
$1\!+\!B(\deltaz_i)/S(\deltaz_i)$ is applied to each 
$\deltaz_i$ bin considered in the fit.

After applying the above correction to the dilepton sample,
we measure
$A_T=(0.5\pm1.2)\,\%$ 
from a fit to the distribution of the charge
asymmetry as a function of \deltat\ (see Fig.~\ref{atresultsWbckg}).

\subsection{Systematic uncertainties}
The detection charge asymmetry
is partially corrected by applying an event-by-event weight. 
We assign the residual asymmetries measured
with the single lepton samples (see Sec.~\ref{section:single}), $\pm0.30\%$ 
for the electrons 
and $\pm0.35\%$ for the muons,
as the systematic errors due to charge asymmetry in detection efficiencies. 
After taking into account the dilution factor, 
the total systematic error related to the charge asymmetry in the detection 
efficiencies
are $\pm0.5\%$ and $\pm0.6\%$ for electrons and muons, respectively.

The charge asymmetry of $(-5 \pm 10)\%$ 
measured with the off-resonance data 
leads to a $\pm0.7\%$ uncertainty on the $A_T$ measurement,
determined from the statistical error $\pm 10\%$. 
In a similar way, the charge asymmetry of
$(1.2 \pm 1.4)\%$,
obtained with the events satisfying $\deltaz< 100\mum$,
results in a $\pm 0.9\%$ uncertainty 
on $A_T$. If we assumed  \CP invariance in the decays producing the cascade 
leptons, this uncertainty would vanish.

The other systematic uncertainties 
are due to the background dilution correction.
This correction is measured with the data  from the full dilepton sample with the 
value of \deltamd fixed  to the world average value~\cite{PDG}. The uncertainty on the 
ratio $B/S$ leads to
a $\pm3\%$ multiplicative error on $A_T$, which is negligible.

All sources of systematic uncertainty are listed in 
Table~\ref{sys_table_AT}.
The total systematic uncertainty is
$\pm1.4\%$.  

\begin{table} [htb]
\begin{center}
\begin{tabular}{|l|c|} 
\hline
{\bf Type of systematic error } & \boldmath $\sigma(A_T) (\%)$ \\
\hline
\hline
Electron charge asymmetry in the detection & 0.5 \\ 
Muon charge asymmetry in the detection & 0.6 \\ 
Non-\BB background charge asymmetry & 0.7 \\ 
\BB background charge asymmetry & 0.9 \\ 
\hline\hline
Total & 1.4 \\ 
\hline
\end{tabular}
\caption{Summary of  systematic uncertainties on $A_T$.}
\label{sys_table_AT} 
\end{center}
\end{table}

%
%

\section{Conclusions}

With the 1999--2000 data consisting of 20.7 \invfb on-resonance and 
2.6 \invfb off-resonance,
we have selected 20,381 same-sign dilepton events with a NN approach and a 
rejection of the
cascade leptons from  charged $B$ or unmixed neutral $B$ events
based on the \deltaz information. Charge
asymmetries in the lepton selection are corrected event-by-event 
and the correction method is confirmed 
with the single lepton sample. We have measured a same-sign dilepton
asymmetry of 
$A_T=(0.5\pm1.2\pm1.4)\%$.
From Eq.~\ref{at},
the $A_T$ asymmetry gives a preliminary 
result for the \T and \CP violation 
parameter $\eb$:
$$
\frac{\reb}{1+|\eb|^2}= (1.2\pm2.9\pm3.6) \times 10^{-3}.
$$
This preliminary measurement is the most stringent 
test of \T and \CP violation in 
$\BzBzb$ mixing to date and is consistent with previous
measurements~\cite{others} of $\reb/(1+|\eb|^2)$.
With the formalism~\cite{PhysBook} using the complex parameters $p$ and $q$,
this measurement of $A_T$ gives  $|q/p|=0.998\pm0.006\pm0.007$.

%
%

\section{Acknowledgments}
\label{section:Acknowledgments}

We are grateful for the 
extraordinary contributions of our \pep2\ colleagues in
achieving the excellent luminosity and machine conditions
that have made this work possible.
The collaborating institutions wish to thank 
SLAC for its support and the kind hospitality extended to them. 
This work is supported by the
US Department of Energy
and National Science Foundation, the
Natural Sciences and Engineering Research Council (Canada),
Institute of High Energy Physics (China), the
Commissariat \`a l'Energie Atomique and
Institut National de Physique Nucl\'eaire et de Physique des Particules
(France), the
Bundesministerium f\"ur Bildung und Forschung
(Germany), the
Istituto Nazionale di Fisica Nucleare (Italy),
the Research Council of Norway, the
Ministry of Science and Technology of the Russian Federation, and the
Particle Physics and Astronomy Research Council (United Kingdom). 
Individuals have received support from the Swiss 
National Science Foundation, the A. P. Sloan Foundation, 
the Research Corporation,
and the Alexander von Humboldt Foundation.


\begin{thebibliography}{9}
\bibitem{CP64} J.\ H.\ Christenson, J.\ W.\ Cronin, V.\ L.\ Fitch 
and R.\ Turlay, \jprl{13}, 138 (1964).
\bibitem{CPLEAR99} CPLEAR Collaboration, A.\ Apostolakis et al.,
\jpl{B456}, 297 (1999).
\bibitem{BB99} M.\ C.\ Ba$\tilde{\rm n}$uls and J.\ Bernab\'eu,
\np{B590}, 19 (2000).
\bibitem{PhysBook} ``The \babar\ Physics Book,''
ed. P.\ F.\ Harrison and H.\ R.\ Quinn, SLAC-R-504 (1998).
\bibitem{SM}T.\ Altomari, L.\ Wolfenstein and J.\ D.\ Bjorken, \jprd{\bf 37}, 1860 (1988);
A.\ Acuto and D.\ Cocolicchio, \jprd{\bf 47}, 3945 (1993). 
\bibitem{BaBarNIM01} \babar\ Collaboration, B.\ Aubert et al.,
SLAC-PUB-8569, \hepex{0105044}, to appear in Nucl. Instrum. and Methods.
\bibitem{FW} G.\ C.\ Fox and S.\ Wolfram, \jprl{\bf 41}, 1581 (1978).
\bibitem{PDG} Particle Data Group, D.\ E.\ Groom et al., \epjc{15}, 1 (2000).
\bibitem{others} CDF Collaboration, F.\ Abe et al., \jprd{55}, 2546 (1997);
OPAL Collaboration, G.\ Abbiendi et al., \epjc{12}, 609 (1999);
ALEPH Collaboration, R.\ Barate et al., \epjc{20}, 431 (2001);
CLEO Collaboration, D.\ E.\ Jaffe et al., \jprl{86}, 5000 (2001).


\end{thebibliography}
\end{document}